\documentclass[prl,twocolumn,preprintnumbers,amsmath,amssymb]{revtex4}
\usepackage{graphicx}

\begin{document}


\title{ Quantum magnetoresistive $(hc/2e)/m$ periodic oscillations in a
superconducting ring}


\author{V.\,I.~Kuznetsov}
\email{kvi@iptm.ru}
\author{O.\,V.~Trofimov}
\affiliation{Institute of Microelectronics Technology and High
Purity Materials, Russian Academy of Sciences, Chernogolovka,
Moscow Region 142432, Russia}

\date{\today}

\begin{abstract}
It was experimentally found that quantum magnetoresistive $hc/2e$
periodic oscillations of the Little-Parks type in a
superconducting mesoscopic ring with decreasing temperature and
increasing applied dc current are modified to the sum of harmonic
$(hc/2e)/m$ periodic oscillations. Multiple Andreev reflection can
be a possible cause of this effect.
\end{abstract}

\maketitle

\section{INTRODUCTION}
In order that to describe a multiply connected superconductor
pierced by a magnetic flux $\Phi$, F. London proposed \cite{c1}
the concept of the superconducting fluxoid $\Phi^{\ast}$ defined
as $\Phi^{\ast}=\Phi+(4\pi/c)\oint\lambda_{L}^{2}$
\textbf{J}$_{s}d$\textbf{s}$\,=\Phi+(m^{\ast}c/e^{\ast})\oint$
\textbf{v}$_{s}d$\textbf{s} (here $c$ is the velocity of light,
$\lambda_{L}$ is the London penetration depth of the magnetic
field, \textbf{J}$_{s}$ is the density of the circulating
superconducting current, \textbf{v}$_{s}$ is the superconducting
velocity, $m^{\ast}$ and $e^{\ast}$ are the effective mass and
effective electric charge of the superconducting pair,
respectively). A remarkable property of superconductivity is that
the superconducting fluxoid $\Phi^{\ast}$ is quantized, that is,
$\Phi^{\ast}=n(hc/e^{\ast})=n\Phi_{0}$ (where $n$ is an integer,
$h$ is the Planck's constant, $e^{\ast}=2e$, $e$ is the electron
charge, $\Phi_{0}=hc/e^{\ast}=hc/2e$ is the superconducting
magnetic flux quantum). Moreover, the found experimental value
$e^{\ast}=2e$ confirms that the effective charge of the
superconductor is $2e$. In particular, quantization fluxoid leads
to quantum oscillations of superconducting circulating current and
the superconducting critical temperature $T_{c}$ (the Little-Parks
effect \cite{c2}) depending on the axial magnetic field $B$ in a
thin-walled superconducting cylinder pierced by the flux and
biased with a very low direct current $I_{dc}$, at temperatures
$T$ very close to $T_{c}$. The periods of these oscillations
correspond to the superconducting magnetic flux quantum
$\Phi_{0}=hc/2e$ through the average cross-sectional area of the
cylinder $S$.

The Little-Parks effect was observed in cylinders with a small
radius $r \approx \xi(T)/2$ (where $\xi(T)$ is a
temperature-dependent superconducting coherence length) under
conditions very close to the equilibrium state. Under conditions
close to a nonequilibrium state (at low temperatures and high
currents), some violations in the $hc/2e$ periodicity of quantum
magnetoresistive oscillations as a function of the axial magnetic
field were found in inhomogeneous doubly connected structures with
a small cross-sectional area: superconducting loops \cite{c3,c4},
Au$_{0.7}$In$_{0.3}$ cylinders \cite{c5} and hybrid metal-normal
Ag rings with superconducting Al mirrors \cite{c6}. Anomalous
negative magnetoresistance (NMR) and the absence of quantum
$hc/2e$ periodic $R(B)$ oscillations were detected in low fields
at low temperatures  and high currents \cite{c3}. Deviations from
the $hc/2e$ periodicity of Little-Parks oscillations in low fields
were found in \cite{c4}. In addition, the NMR and $hc/4e$ periodic
$R(B)$ oscillations were observed in superconducting cylinders
\cite{c5} and hybrid SNS structures \cite{c6} at low temperatures.
The period of $hc/4e$ was explained by the appearance of a new
minimum of the free energy at $\Phi$\,$=$\,$\Phi_{0}(n+1/2)$ due
to the formation of a $\pi$-junction in the Au$_{0.7}$In$_{0.3}$
cylinder \cite{c5} and multiple Andreev reflection \cite{c6}.

Quantum oscillations in superconducting loops of a larger area
under clearly nonequilibrium conditions (high currents and $T$
below $T_{c}$) are not practically studied. In rings with larger
radii $r \approx 2\xi(T)$, the amplitude of quantum oscillations
should be expected to be low due to the lower circulating
superconducting current and due to the fact that not all the ring
can switch from the superconducting state to the normal state and
back with a change in $B$.

The study of quantum magnetoresistive oscillations in
superconducting mesoscopic rings without weak links (or without
tunnel junctions) is challenging, since such rings can work as a
highly sensitive superconducting quantum interference device
(SQUID) \cite{c7} and as a highly efficient magnetic-dependent ac
voltage rectifier (if the ring has circular asymmetry)
\cite{c8,c9}.

\begin{figure}
\includegraphics[width=1\linewidth]{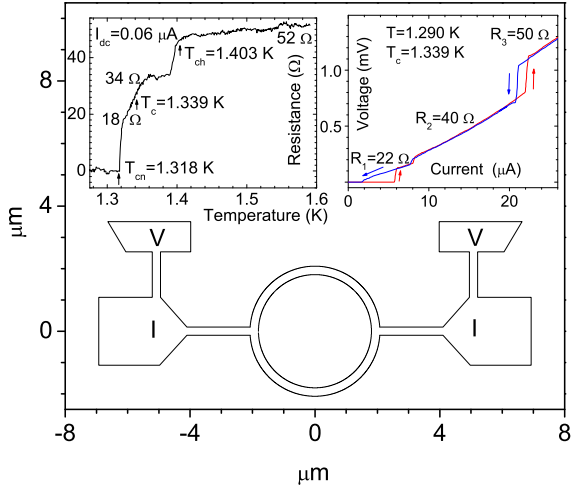}
\caption{\label{f1} (Color online) The sketch of the structure.
The letters $I$ and $V$  denote the current and potential wires.
The left inset: Resistive $R(T)$ transition at
$I_{dc}$\,=\,$0.06$\,$\mu$A. The right inset: $V(I)$ function at
$T$\,=\,$1.290$\,K in the zero field; the arrows indicate the
direction of the dc sweep.}
\end{figure}

It was found that periods of quantum magnetoresistive oscillations
under strongly nonequilibrium conditions (at a high direct current
$I_{dc}$ and $T$  below $T_{c}$)) can be $m$  times lower than the
usual value and correspond to $\Phi_{0}/m$\,$=$\,$(hc/2e)/m$ ($m$
is an integer) through the effective area of the superconducting
ring with a larger radius ($r=2$\,$\mu$m\,$>\xi(T)$).

\section{SAMPLES AND EXPERIMENTAL PROCEDURE}
$V(B)$ voltage was measured at different direct currents
$I_{dc}>I_{c}$ at temperatures $T$ slightly below $T_{c}$ in
superconducting mesoscopic structures (with close and different
geometry), pierced by a magnetic flux. The structure under study
(Fig.\,\ref{f1}) was obtained by thermal evaporation of an
aluminum film with a thickness of $d=51$\,nm on a silicon
substrate using the lift-off process of electron-beam lithography.
The central region of the structure consists of a ring having a
wall thickness $w_{r}$\,=\,$0.27$\,$\mu$m and an average radius
$r_{m}$\,=\,$1.94$\,$\mu$m, narrow current $I$ wires with a width
$w_{n}$\,=\,$0.27$\,$\mu$m, wide current wires with a width
$w_{w}$\,=\,$2$\,$\mu$m and potential $V$  leads. Voltage $V$ was
measured in the region including the ring, narrow $I$ wires and
parts of wide $I$ wires (Fig.\,\ref{f1}).

The structure had the parameters: resistance
$R_{4.2K}=52.7$\,$\Omega$ at $T=4.2$\,K, resistance per square of
film thickness $R_{sq}=\rho /d=1.97\,\Omega$, the ratio of
resistances at $T=300$\,K and 4.2\,K is equal to
$R_{300K}/R_{4.2K}=1.8$. The mean free path of quasiparticles
$l=10$\,nm was found from the refined theoretical \cite{c10}
relation $\rho l=5.1 \times 10^{-16}$\,$\Omega$\,$m^{2}$, where
$\rho$ is the resistivity of the wire. The structure is a dirty
superconductor, since $l<<\xi_{0}$ (where $\xi_{0}=1.6$\,$\mu$m is
the superconducting coherence length of pure aluminum at
$T=0$\,K). Near $T_{c}$ for the dirty case \cite{c11},
$\xi(T)=\xi(0)(1-T/T_{c})^{-1/2}$ (where
$\xi(0)=0.85(\xi_{0}l)^{1/2}=0.11$\,$\mu$m). The condition of
quasi-one-dimensional superconductivity ($w_{n}$, $w_{r}<2\xi(T)$)
is fulfilled near $T_{c}$. The nonequilibrium diffusion length of
quasiparticles \cite{c3,c12} $\lambda_{Q}(T,I_{dc},B)=6-9$\,$\mu$m
near $T_{c}$. The structure has a length $L=10$\,$\mu$m (the
distance between $V$ leads), satisfying the condition
$\xi(T)<<L<2\lambda_{Q}(T,I_{dc},B)$.

\section{RESULTS AND DISCUSSION}
The resistive $R(T)$ transition of the structure from the normal
(N) state to the superconducting (S) state is recorded at a direct
current $I_{dc}$\,=\,$0.06$\,$\mu$A (the left inset of
Fig.\,\ref{f1}). The transition is rather stretched, which
indicates the heterogeneity of the structure. The beginning of a
sharp drop in $R(T)$ occurs at $T_{ch}$\,=\,$1.403$\,K and the
resistance disappears at $T_{cn}=1.318$\,K. The superconducting
critical temperature $T_{c}$\,=\,$1.339$\,K is determined by the
middle of the $R(T)$ transition. The contribution of the ring,
expected from the geometry, to the total resistance of the
structure is 23\,$\Omega$. We assume that the upper segment of the
$R(T)$ transition (from 52 to 18\,$\Omega$) and the lower segment
of $R(T)$ (from 18 to 0\,$\Omega$) correspond to the NS
transitions of the current wires and the ring, respectively. The
function $V(I)$, recorded at $T$\,=\,$1.290$\,K, demonstrates
phase current separation into sections with different resistance
(the right inset of Fig.\,\ref{f1}). The initial segment of the
$V(I)$ function at low currents, including a nearly linear section
with a resistance of 22\,$\Omega$, characterizes SN (NS)
transitions in the ring.

\begin{figure}
\includegraphics[width=1\linewidth]{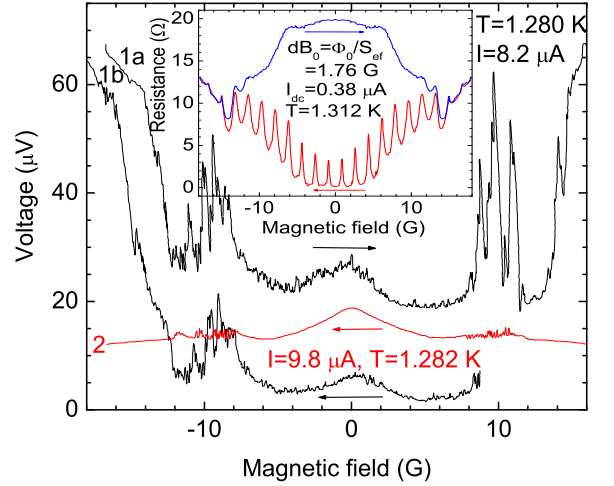}
\caption{\label{f2} (Color online) $V(B)$ functions measured at
$I_{dc}=8.2$\,$\mu$A and $T=1.280$\,K (curves\,1a,\,1b) and at
$I_{dc}=9.8$\,$\mu$A and $T=1.282$\,K (curve\,2). The arrows
indicate the field sweep direction. The inset shows $V(B)$
function at $I_{dc}=0.38$\,$\mu$A and $T=1.312$\,K.}
\end{figure}

The magnetic field changed from a conditionally negative value of
-\,20\,G to a conditionally positive value of +\,20\,G and back
when measuring the $V(B)$ functions. In the field interval
-\,20,\,+20\,G, only a part of the structure corresponding to the
ring, switched into a resistive state with a resistance not
exceeding $20$\,$\Omega$ (Fig.\,\ref{f2} and the inset). The
experimental $V(B)$ functions show an anomalous hysteresis
depending on the direction of the field sweep (Fig.\,\ref{f2} and
the inset). The ring has two states: a more dissipative state (the
curve\,1a of the Fig.\,\ref{f2} and the upper curve of the inset
of Fig.\,\ref{f2}) and a less dissipative state (the curves\,1b,
\,2 of the Fig.\,\ref{f2} and the lower curve of the inset of
Fig.\,\ref{f2}). The causes of two states will be analyzed
elsewhere. The considerable difference between two states is seen
in the inset of Fig.\,\ref{f2}.

The inset of Fig.\,\ref{f2} shows the $R(B)$ resistance as a
function of the field, measured at a low current
$I_{dc}$\,=\,$0.06$\,$\mu$A and $T$$\,=\,$1.312\,K, very close to
the bottom of the NS transition. The feature of the upper curve
(the inset of Fig.\,\ref{f2}) is the anomalous negative
magnetoresistance (NMR), reaching a maximum at $B$\,=\,$0$, equal
to the ring resistance in the normal state. The anomalous
dissipative state arises due to the thermodynamic fluctuations of
the superconducting order parameter, leading to the formation of a
phase slip center (PSC) \cite{c12} in the ring, despite the low
current. Quantum magnetoresistive $hc/2e$ periodic oscillations of
the Little-Parks type are visible on the lower curve (the inset of
Fig.\,\ref{f2}). The fundamental magnetic-field period of the
oscillations is
$dB_{0}$\,=\,$\Phi_{0}/S_{eff}$\,=\,$(hc/2e)/S_{eff}$\,=\,$1.76$\,G
and corresponds to the superconducting magnetic flux quantum
$\Phi_{0}$\,=\,$hc/2e$  through the effective ring area $S_{eff}$.
$S_{eff}$ almost coincides with the mean geometric area of the
ring. The fundamental frequency
$f_{0}$\,=\,$dB_{0}^{-1}$\,=\,$0.569$\,G$^{-1}$ has the meaning of
the reciprocal of the fundamental oscillation period $dB_{0}$.

The $V(B)$ functions recorded at lower $T$\,=\,$1.280-1.284$\,K
and high currents $I_{dc}$\,=\,$7.5-11$\,$\mu$A also have a
hysteresis decreasing with increasing $I_{dc}$. In addition,  the
curves\,1a, \,1b, and 2 of the Fig.\,\ref{f2} show anomalous
negative magnetoresistance in two field intervals: fields close to
zero and low fields. Figure\,\ref{f2} shows two of these unusual
$V(B)$ functions measured at $I_{dc}$\,=\,$8.2$\,$\mu$A and
$T=1.280$\,K (curves\,1a,\,1b) and at $I_{dc}= 9.8$\,$\mu$A and
$T$\,=\,$1.282$\,K (curve\,2). The right segment of curve\,1b and
the upper curve, close to curve\,2, recorded for the other
direction of the field sweep, are not shown in Fig.\,\ref{f2}.

\begin{figure}
\includegraphics[width=1\linewidth]{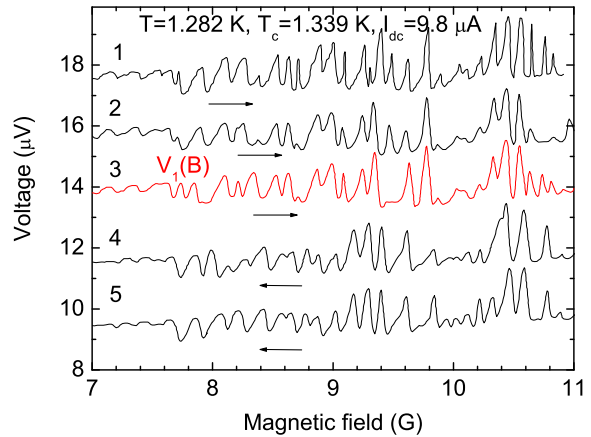}
\caption{\label{f3} (Color online) $V(B)$ functions measured many
times at $I_{dc}=9.8$\,$\mu$A and $T=1.282$\,K in the fields
$B=7-11$\,G. The arrows indicate the direction of the field sweep.
The $V(B)$ functions, except $V_{1}(B)$ function (curve\,3) are
shifted vertically relative to the neighboring function by
2\,$\mu$V.}
\end{figure}

Negative magnetoresistance appeared in a threshold manner at a
certain current $I_{dc}>I_{r}$. Here $I_{r}$  is the retrapping
superconducting critical current at which $V(I)$ on the structure
disappears with decreasing $I_{dc}$. In addition, unusual $V(B)$
oscillations were found against the background of the NMR near the
zero field at currents $I_{dc}$\,=\,$1-3$\,$\mu$A (not shown here)
and low fields (6-12\,G) at high currents
$I_{dc}$\,=\,$7.5-11$\,$\mu$A (Fig.\,\ref{f2}). Figure\,\ref{f3}
shows the right segment of curve \,2 (Fig.\,\ref{f2}) measured
several times and demonstrating unusual oscillations. These
oscillations are not noise, they are almost reproducible when
re-recording and differ slightly depending on the direction of the
field sweep.

The Fourier spectrum of oscillations is usually calculated for a
detailed analysis. The Fourier spectrum of any $x(t)$
oscillations, existing in a limited interval from $t_{1}$  to
$t_{2}$, contains, besides the physical frequencies, fictitious
frequencies: zero frequency and frequencies
$f_{k}$\,=\,$k$\,$dt_{1,2}^{-1}$ (where $k$ is an integer,
$dt_{1,2}$\,=\,$t_{2}-t_{1}$ - interval length). Moreover, the
physical frequencies can be shifted by the value of
$k$\,$dt_{1,2}^{-1}$. Fictitious low frequencies were observed in
the spectra of the $V(B)$ functions in \cite{c9,c13}. It was found
that for $V(B)$ oscillations (Fig.\,\ref{f3}) the fundamental
frequency $f_{0}$ is close in order of value to $dB_{1,2}^{-1}$
(here $dB_{1,2}$\,=\,$B_{2}-B_{1}$ is the length of the
oscillation existence interval), therefore distortions are
expected in the Fourier spectrum.

\begin{figure}
\includegraphics[width=1\linewidth]{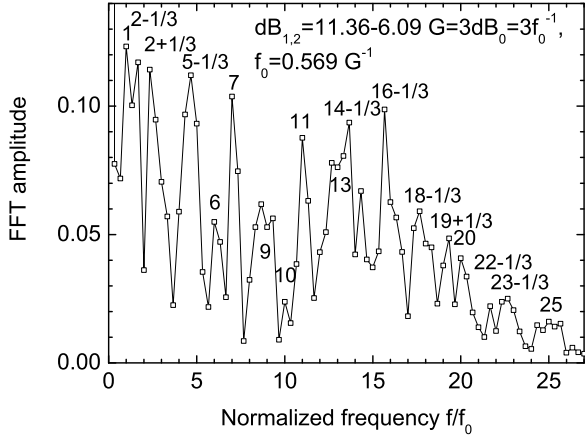}
\caption{\label{f4} The FFT spectrum of the $V_{1}(B)$ function
(the curve\,3 of Fig.\,\ref{f3}), taken in the interval of fields
from $6.09$ to $11.36$\,G. The condition
$dB_{1,2}=11.36-6.09=3dB_{0}=3f_{0}^{-1}$ is satisfied.}
\end{figure}

\begin{figure}
\includegraphics[width=1\linewidth]{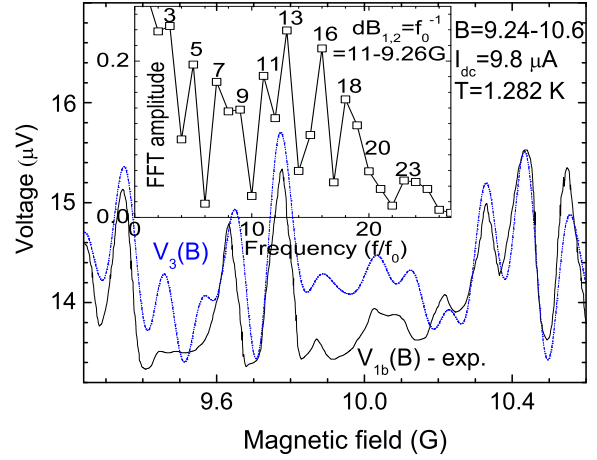}
\caption{\label{f6} (Color online) $V_{1b}(B)$ function (solid
line) is the part of the experimental $V_{1}(B)$ function (the
curve\,3 of Fig.\,\ref{f3}), shown in fields $9.24$\,-\,$10.6$\,G;
$V_{3}(B)$ function (dash-dotted line) is the qualitative fitting
of the $V_{1b}(B)$ function. The inset: Fourier amplitude as a
function of $f/f_{0}$, obtained from the $V_{1b}(B)$ function in
fields $9.26$\,-\,$11.02$\,G. The condition
$dB_{1,2}=11.02-9.26=dB_{0}=f_{0}^{-1}$ is fulfilled.}
\end{figure}

In order to better see the distortions in the spectra, we plotted
the Fourier transform (Figs.\,\ref{f4}\,-\,\ref{f6}) of the $V(B)$
oscillations with the given interval lengths
$dB_{1,2}=j$\,$dB_{0}=j$\,$f_{0}^{-1}$ (where $j=3$\, and \,1).
The zero and fictitious frequencies
$f_{k,j}=k$\,$dB_{1,2}^{-1}=(k/j)dB_{0}^{-1}=(k/j)f_{0}^{-1}$ were
expected to find in the spectrum, besides the physical
frequencies. In addition, physical frequencies were expected to
shift by $f_{k,j}$.

Fast Fourier transforms (FFTs) were obtained using 16384 evenly
distributed points in given intervals of fields. So the condition
$dB_{1,2}=B_{2}-B_{1}=11.36-6.09=3dB_{0}=3f_{0}^{-1}$ is fulfilled
for the FFT spectrum of the $V_{1}(B)$ function (the curve\,3 of
Fig.\,\ref{f3}) taken in the field interval from 6.09 to 11.36\,G.
This spectrum (Fig.\,\ref{f4}) contains the fundamental frequency
$f_{0}=dB_{0}^{-1}=S_{ef} /\Phi_{0}=0.569$\,G$^{-1}$. The values
of $f_{0}$ found from the period of the Little-Parks type
oscillations (the inset of Fig.\,\ref{f2}) and the spectrum
(Fig.\,\ref{f4}) coincide. In addition to $f_{0}$, the spectrum
contains many higher harmonics of the fundamental frequency
$f_{0}$, defined as $f_{m}=mf_{0}$ (where $m=2-20$). Some
frequency peaks (Fig.\,\ref{f4}) are shifted by 1/3, since the
condition $dB_{1,2}=11.36-6.09=3dB_{0}=3f_{0}^{-1}$ is specified.
All FFT spectra, including this spectrum (Fig.\,\ref{f4}), contain
a fictitious zero frequency. The contributions of the fundamental
frequency and many higher harmonics to the spectrum are close.
This indicates the presence of various fractional
($\Phi_{0}/m=(hc/2e)/m$) periods of $V_{1}(B)$ oscillations that
are not a consequence of the inharmonicity of the $hc/2e$
oscillations. Some fractional $(hc/2e)/m$ magnetic flux periods,
corresponding to fractional $dB_{0}/m$ magnetic field periods, are
clearly distinguishable on the $V_{1}(B)$ function (the curve\,3
of Fig.\,\ref{f3}). Inharmonicity of oscillations is believed to
make a very low contribution to the higher harmonics of the
fundamental frequency $f_{0}$.

\begin{figure}
\includegraphics[width=1\linewidth]{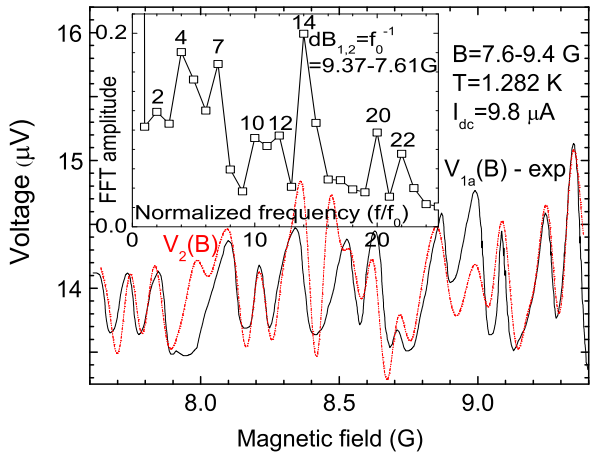}
\caption{\label{f5} (Color online) The $V_{1a}(B)$ function (solid
line) is part of the experimental $V_{1}(B)$ function (the
curve\,3 of Fig.\,\ref{f3}), shown in fields $7.61$\,-\,$9.37$\,G;
the $V_{2}(B)$ function (dash - dotted line) - qualitative fitting
of the $V_{1a}(B)$ function. The inset: Fourier amplitude as a
function of the normalized frequency $f/f_{0}$, obtained from the
$V_{1a}(B)$ function in fields $7.61$\,-\,$9.37$\,G. The condition
$dB_{1,2}$\,=\,$9.37$\,-\,$7.61$\,=\,$dB_{0}$\,=\,$f_{0}^{-1}$ is
satisfied.}
\end{figure}
For a detailed analysis, the experimental $V_{1}(B)$ function (the
curve\,3 of Fig.\,\ref{f3}) is divided into two functions
$V_{1a}(B)$ in fields 7.6-9.4\,G (Fig.\,\ref{f5}) and $V_{1b}(B)$
in fields 9.24-10.6\,G (Fig.\,\ref{f6}). The FFT spectra of both
$V_{1a}(B)$ and $V_{1b}(B)$ functions are calculated in two field
intervals of 7.61-9.37\,G and 9.26-11.02\,G, respectively (the
insets of Figs.\,\ref{f5}\, and \,\ref{f6}). Unlike the spectrum
(Fig.\,\ref{f4}), the spectra (Figs.\,\ref{f5}\, and \,\ref{f6})
do not contain peaks shifted in frequency, since the conditions
$dB_{1,2}=B_{2}-B_{1}=9.37-7.61=dB_{0}=f_{0}^{-1}$ and
$dB_{1,2}=11.02-9.26=dB_{0}=f_{0}^{-1}$ were specified. Certain
numbers of $m$ dominate. The apparent absence of frequencies with
other values of $m$ in the spectra, including $m=1$, is due to the
low spectral resolution broadening the frequency peaks.

Only in order to qualitatively show that the $V_{1}(B)$ function
really has a certain set of different oscillation periods, the
different  sections of $V_{1}(B)$ were approximated by fitting
functions. The fitting has nothing to do with the theoretical
description of the $V_{1}(B)$ oscillations. The expression
$s+p\Sigma ^{k}(a_{k} sin(2\pi mf_{0}B+\varphi_{k}))$, was used
for the fitting, consisting of a constant shift $s$ and the
multiplication of the coefficient $p\approx 1$ to the sum of
sinusoidal oscillations with different amplitudes $a_{k}$,
frequencies $mf_{0}$ and phases $\varphi_{k}$ that are multiples
of $\pi /4$. The index $m$, corresponding to the harmonic number,
took some integer values. The results of the spectra were taken
into account (Figs.\,\ref{f5}\, and \,\ref{f6}); therefore,
$a_{k}$ are close to the Fourier amplitudes obtained from the
spectra. Other options are available for fitting of the $V_{1}(B)$
function.

Two fitting functions $V_{2}(B)$ and $V_{3}(B)$ are used,
respectively, for a qualitative description of $V_{1a}(B)$
(Fig.\,\ref{f5}) and $V_{1b}(B)$ (Fig.\,\ref{f6}) functions that
are parts of the experimental $V_{1}(B)$ function. The prevalence
in the spectra (the insets of Figs.\,\ref{f5}\, and \,\ref{f6}) of
certain frequencies $f_{m} $\,=\,$mf_{0}$ (where
$m$\,=\,$2$,\,4,\,7,\,10,\,12,\,14,\,20,\,22 for $V_{2}(B)$ and
$m$\,=\,$3$,\,5,\,7,\,9,\,11,\,13,\,16,\,18 for $V_{3}(B)$) is
taken into account. The functions of $V_{2}(B)$ and $V_{3}(B)$ are
given below.

$V_{2}(B)=14.05+1.3(0.12sin(2\pi2 f_{0}B-\pi /2)+$

$+0.18sin(2\pi 4f_{0}B)+0.17sin(2\pi 7 f_{0}B+\pi /4)+$

$+0.1sin(2\pi 10 f_{0}B-\pi /4)+0.1sin(2\pi 12 f_{0}B+\pi /2)+$

$+0.2sin(2\pi 14 f_{0}B-\pi /2)+0.1sin(2\pi 20 f_{0}B)+$

$+0.08sin(2\pi 22 f_{0}B+3\pi /4))$.

$V_{3}(B)=14.3+1.25(0.245sin(2\pi 3 f_{0}B+\pi)+$

$+0.2sin(2\pi 5 f_{0}B-3\pi /4)+0.17sin(2\pi 7 f_{0}B+\pi /4)+$

$+0.14sin(2\pi 9 f_{0}B+\pi /4)+0.18sin(2\pi 11 f_{0}B)+$

$+0.24sin(2\pi 13f_{0}B)+0.22sin(2\pi 16 f_{0}B+\pi /2)+$

$+0.15sin(2\pi 18 f_{0}B+3\pi /4)) $.

\begin{figure}
\includegraphics[width=1\linewidth]{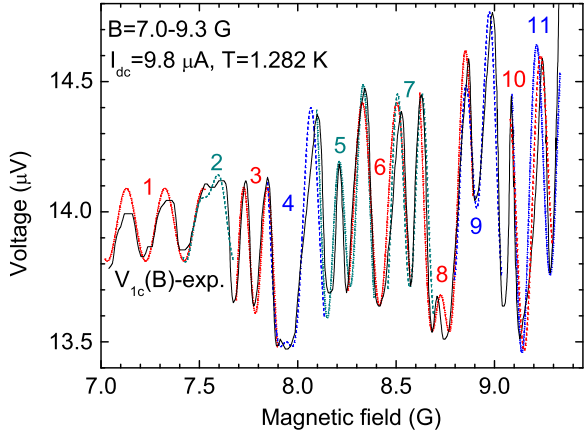}
\caption{\label{f7} (Color online) $V_{1c}(B)$ function (solid
line) is the part of the $V_{1}(B)$ function (the curve\,3 of
Fig.\,\ref{f3}), shown in fields $7.0$\,-\,$8.3$\,G. The
curves\,1,\,3,\,5,\,6,\,8,\,11 (dash-dotted lines) and
2,\,4,\,7,\,9,\,10 (dashed lines) are approximations of individual
sections of the $V_{1c}(B)$ function with a set of fitting
functions $F_{k}(B)$.}
\end{figure}

\begin{figure}
\includegraphics[width=1\linewidth]{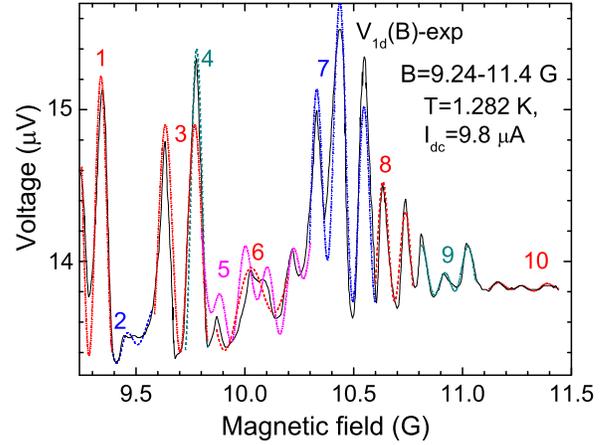}
\caption{\label{f8} (Color online) $V_{1d}(B)$ function (solid
line) is the part of the $V_{1}(B)$ function (the curve\,3 of
Fig.\,\ref{f3}), shown in fields $9.24$\,-\,$11.4$\,G. The
curves\,1,\,3,\,5,\,7,\,9,\,10 (dash-dotted lines) and
2,\,4,\,6,\,8 (dashed lines)  are approximations of individual
sections of the $V_{1d}(B)$ function with a set of fitting
functions $A_{k}(B)$.}
\end{figure}

The $V_{1c}(B)$ and $V_{1d}(B)$ functions that are parts of the
experimental $V_{1}(B)$ function (the curve\,3 of Fig.\,\ref{f3})
are shown in Figures \ref{f7} and \ref{f8} with solid lines,
approximations of short sections of both functions are shown with
the broken lines. The short sections with the numbers 1\,-\,11 of
the $V_{1c}(B)$ function (Fig.\,\ref{f7}) and the sections
1\,-\,10 of the $V_{1d}(B)$ function (Fig.\,\ref{f8}) are fitted
by a single sinusoidal oscillation or a sum of several
oscillations with different amplitudes, frequencies
$f_{m}$\,=\,$mf_{0}$ and phases.

Sets of fitting functions $F_{k}(B)$ (Fig.\,\ref{f7}) are
described for the curves: 1\,(in the fields 7.0-7.5\,G),
2\,(7.41-7.67\,G), 3\,(7.68-7.89\,G), 4\,(7.84-8.13\,G),
5\,(8.09-8.38\,G), 6\,(8.255-8.57\,G), 7\,(8.45-8.69\,G),
8\,(8.62-8.89\,G), 9\,(8.79-9.03\,G), 10\,(9.08-9.3\,G),
11\,(9.08-9.3\,G) with the following expressions:

$F_{1}(B)=13.95+0.14sin(2\pi9f_{0}B-\pi/2)$;

$F_{2}(B)=14+0.14sin(2\pi7f_{0}B+\pi/4)+$

$+0.06sin(2\pi14f_{0}B-3\pi/4)$;

$F_{3}(B)=13.8+0.1sin(2\pi7f_{0}B+\pi/2)+ $

$+0.29sin(2\pi14f_{0}B-\pi/2)$;

$F_{4}(B)=13.8+0.45sin(2\pi7f_{0}B+\pi/4)+ $

$+0.15sin(2\pi14 f_{0}B)$;

$F_{5}(B)=14+0.16sin(2\pi7f_{0}B+\pi/4)+ $

$+0.34sin(2\pi14f_{0}B-\pi/4)$;

$F_{6}(B)=14.03+0.39sin(2\pi10f_{0}B-\pi/4)$;

$F_{7}(B)=14.0+0.1sin(2\pi4f_{0}B-\pi/2)+ $

$+0.39sin(2\pi14f_{0}B-\pi)$;

$F_{8}(B)=13.9+0.47sin(2\pi7f_{0}B)+ $

$+0.25sin(2\pi14f_{0}B-\pi/2)$;

$F_{9}(B)=14.25+0.2sin(2\pi7f_{0}B-3\pi/4)+ $

$+0.37sin(2\pi14f_{0}B-\pi/2)$;

$F_{10}(B)=14.15+0.17sin(2\pi7f_{0}B+\pi/2)+ $

$+0.56sin(2\pi12f_{0}B+3\pi/4)$;

$F_{11}(B)=14.05+0.25sin(2\pi7f_{0}B+3\pi/4)+ $

$+0.36sin(2\pi12f_{0}B+\pi/2)$,

respectively. It can be seen that the oscillation frequencies used
to approximate the short sections 1-11 of the $V_{1c}(B)$ function
(Fig.\,\ref{f7}) are as follows: $f=9f_{0}$ (section 1); $7f_{0}$
and 14$f_{0}$ (sections 2, 3, 4, 5, 8, 9); 10$f_{0}$ (section 6);
4$f_{0}$, and 14$f_{0}$ (section 7); $7f_{0}$, and $12f_{0}$
(sections 10, and 11).

The fitting functions $A_{k}(B)$ (Fig.\,\ref{f8}) are described
for the curves: 1 ($B=9.25-9.4$\,G), 2 (9.38-9.58\,G), 3
(9.58-9.83\,G), 4 (9.73-9.83\,G), 5 (9.8-10.3\,G), 6
(9.9-10.2\,G), 7 (10.3-10.6\,G), 8 (10.6-10.77\,G), 9
(10.81-11.06\,G), 10 (11.12-11.44\,G) with the following
expressions:

$A_{1}(B)=14.3+0.92sin(2\pi16f_{0}B+\pi /2)$;

$A_{2}(B)=13.565+0.135sin(2\pi5f_{0}B-\pi/4)+$

$+0.15sin(2\pi16f_{0}B)+0.1sin(2\pi18f_{0}B-3\pi/4)$;

$A_{3}(B)=14.15+0.75sin(2\pi13f_{0}B)$;

$A_{4}(B)=14.4+sin(2\pi16f_{0}B+\pi/2)$;

$A_{5}(B)=13.9+0.85(0.24sin(2\pi3f_{0}B-3\pi/4)+$

$+0.19sin(2\pi5f_{0}B-3\pi/4)+0.17sin(2\pi7f_{0}B+3\pi/4)+$

$+0.22sin(2\pi16f_{0}B+\pi/2))$;

$A_{6}(B)=13.78+1.1(0.2sin(2\pi5f_{0}B+3\pi/4)+$

$+0.3sin(2\pi7f_{0}B+3\pi/4))$;

$A_{7}(B)=14.6+1.2(0.3sin(2\pi7f_{0}B-\pi/2)+$

$+0.65sin(2\pi16f_{0}B+\pi/2))$;

$A_{8}(B)=14.05+1.1(0.16sin(2\pi11f_{0}B+\pi)+$

$+0.4sin(2\pi16f_{0}B+\pi))$;

$A_{9}(B)=13.9+1.2(0.077sin(2\pi9f_{0}B-\pi/4)$+

$+0.097sin(2\pi16f_{0}B-\pi/4))$;

$A_{10}(B)=13.83+1.2(0.009sin(2\pi7f_{0}B-\pi/2)+$

$+0.015sin(2\pi16f_{0}B-3\pi/4))$,

respectively. The oscillation frequencies used to fit the sections
1\,-\,10 of the $V_{1d}(B)$ function (Fig.\,\ref{f8}) are as
follows: $16f_{0}$ (section 1); $5f_{0}$, $16f_{0}$ and $18f_{0}$
(section 2); $13f_{0}$ (section 3); $16f_{0}$ (section 4);
$3f_{0}$, $5f_{0}$, $7f_{0}$ and $16f_{0}$ (section 5); $5f_{0}$
and $7f_{0}$ (section 6); $7f_{0}$ and $16f_{0}$ (section 7);
$11f_{0}$ and $16f_{0}$ (section 8); $9f_{0}$ and $16f_{0}$
(section 9); $7f_{0}$ and $16f_{0}$ (section 10).

The detailed analysis, including the Fourier transform of $V(B)$
oscillations (the curve\,2 of Fig.\,\ref{f2} measured at
$I_{dc}$\,=\,$9.8 \mu$A and $T$\,=\,$1.282$\,K in fields 6-12\,G)
has been done. It can be seen from the FFT spectrum
(Fig.\,\ref{f4}) that the contributions of the fundamental
frequency $f_{0}$ and the majority of higher harmonics
$f_{m}$\,=\,$mf_{0}$ ($m$\,=\,$2-20$) to the spectrum are close.
Each of the remaining higher harmonics makes a contribution that
is approximately two times lower than the contribution of $f_{0}$.

In addition, other unusual experimental $V(B)$ oscillations were
measured in low fields of 6-12\,G at $T$\,=\,$1.280-1.284$\,K (not
presented here). The maximum oscillation amplitude decreased from
25\,$\mu$V to 0 with an increase in the current $I_{dc}$ from 7.5
to 11\,$\mu$A. Thus, this amplitude reached 20 and 2\,$\mu$V at
currents of 8.2 and 9.8\,$\mu$A (Fig.\,\ref{f2}), respectively. At
currents $I_{dc}$\,=\,$7.7-8.6$\,$\mu$A, the contributions of
higher harmonics $f_{m}$\,=\,$mf_{0}$ ($m$\,=\,$3-20$) to the
spectra (not presented here) were close, but approximately two
times less than the contributions of frequencies $f_{0}$ and
$2f_{0}$.

Previously, negative magnetoresistance near $B\,=\,0$ was measured
in superconducting quasi-one-dimensional wires with
cross-sectional narrowing and in rings of small radii with
inhomogeneities \cite{c3}. Quantum magnetoresistive oscillations
in the anomalous region of the NMR were not detected in \cite{c3}.
Two regions of negative magnetoresistance (near the zero field and
in the low fields) were found.

A generally accepted mechanism of NMR has not been proposed so
far. The negative magnetoresistance in our structure can be due to
a decrease in the resistance of the nonequilibrium SN boundary
$R_{Q}\approx \lambda_{Q}(T,I_{dc},B)$ with the increasing field
\cite{c3}. Another reason for the NMR can be an increase in the
retrapping superconducting critical current $I_{r}$  and a
decrease in the resistance of the structure (restoring of
superconductivity) at a given current with the increasing field
due to a decrease in the "effective temperature" of hot
quasiparticles in a non-equilibrium region of the structure
\cite{c14}. The "effective temperature" decreases due to an
increase in the diffusion of overheated quasiparticles into
neighboring superconducting banks with a slightly larger value of
the superconducting order parameter $\Delta(T,B)$, when
$\Delta(T,B)$ in superconducting banks decreases (or is completely
suppressed) with the increasing field \cite{c14}.

The non-equilibrium region is the phase slip center or the SNS
junction, in the center of which $\Delta(T,B)$ it periodically
becomes zero and has a time-averaged non-zero value lower than in
neighboring superconducting regions \cite{c12}. Oscillations of
the order parameter give rise to a large quantity of hot
quasiparticles, leading to a strong heating of the nonequilibrium
region. The multiple Andreev reflection (MAR) \cite{c15,c16,c17},
occurring in the phase slip center or the SNS junction at voltages
lower than the superconducting gap, increases the quasiparticle
heating. The heating is enhanced due to the large electron-phonon
relaxation time in the aluminum structure. Superconductivity and
dissipation coexist in the non-equilibrium region \cite{c18}.

Although the theory \cite{c14} is valid only for short samples
with a distance between potential leads
$L$\,=\,$5\xi(T)<2\lambda_{Q}(T,I_{dc},B)$. It can qualitatively
explain the NMR of the experimental $V(B)$ functions in our
structure with an average length $L$\,=\,$20\xi(T)$\,=\,$10$\,
$\mu$m, satisfying the condition
$2\xi(T)<<L<2\lambda_{Q}(T,I_{dc},B)$. We assume that both the NMR
regions in the fields close to the zero and in low fields are due
to the formation of a superconducting barrier for diffusion of hot
quasiparticles at the transition point of a narrow current wire to
the wide current wire.

This superconducting barrier is significantly weakened (or
disappears) in a low field $B_{b}(T,I_{dc})$, dependent on $T$ and
$I_{dc}$. The value $B_{b}(T,I_{dc})$ is close to the third
critical field $B_{c3}(T)$ at low currents. The appearance of
superconductivity along the boundary of a wide current wire
becomes impossible in the fields above $B_{c3}(T)$. The critical
field $B_{c3}(T) \approx 2B_{c2}(T)$ is estimated for the point of
transition of narrow current wires to wide current wires.
Multiplier 2 is taken due to the wedge-shaped transitions
\cite{c19}. For the $V(B)$ function (the inset of Fig.\,\ref{f2}),
the calculated value $B_{c3}(T=1.312K) \approx 2\Phi_{0}/2\pi
\xi(T)^{2}\approx 12$\, G is close to the experimental value of
the field at which the negative magnetoresistance  disappears
($B$\,=\,$14$\,G). For the $V(B)$ function (the curve\,2 of
Fig.\,\ref{f2}), $B_{c3}(T=1.282K) \approx 22-26$\,G was greater
than the field at which the NMR disappears ($B$\,=\,$12$\,G),
since the effect of the large direct current $I_{dc}$ is not taken
into account. We believe that two regions of the NMR are due to
several transitions between the diamagnetic and paramagnetic
states of wide current wires with a change in the field. In the
measurement, the maximum critical field $B_{max}(T)$ is not
reached, at which superconductivity in narrow current wires and
the ring is completely suppressed \cite{c11}.
$B_{max}(T)$\,=\,$77$\,G at the temperature $T$\,=\,$1.282$\,K.

It is common knowledge that quantum magnetoresistive fractional
$(hc/2e)/m$ (with $m>2$) periodic oscillations were not observed.
We believe that oscillations with periods corresponding to
$\Phi_{0}/m=(hc/2e)/m$ (where $m$\,=\,$1-20$) and approximately
equal amplitudes (except amplitude corresponding to $m=1$)
indicate that the superconducting circulating current has
effective charge $e^{\ast}=2em$. This phenomenon can be caused by
the multiple Andreev reflection \cite{c15,c16,c17,c18} occurring
in non-equilibrium regions of the structure (SNS junction or phase
slip center formed in the ring) at voltages
$V(I_{dc})<2\Delta(T,B)/e$. Here
$\Delta(T,B)=\Delta(T)(1-(B/B_{max}(T))^{2})^{1/2}$ is the
superconducting gap in the magnetic field $B$ at $T$ slightly
below $T_{c}$. Where $\Delta(T)=3.07
kT_{c}\Delta(0)(1-T/T_{c})^{1/2}$ is the gap at $B=0$ \cite{c11}.
For the $V(B)$ function (the curve\,2 of Fig.\,\ref{f2}),
$\Delta(T=1.282K, B=6-12 G)=74-73$\,$\mu$V.

It is known that multiple Andreev reflection can be realized in
SNS junctions both in the ballistic case with the mean free path
of quasiparticles $l$ greater than the length $l_{n}$ of the
normal region of the junction \cite{c15} and in the diffusive case
$(l<<l_{n})$ \cite{c16,c17,c18}. In this work, a diffusive the
phase slip center or SNS junction is formed
($l<<l_{n}$\,=\,$2\xi(T))$. To observe a large number of $n$ MAR
in diffusive SNS junctions, it is required that the inelastic
scattering length $l_{in}$ \cite{c16} (in our case, $l_{in}$\,=\,$
2\lambda_{Q}(T,I_{dc},B) $\,=\,$12-18$\,$\mu$m) is much larger
than $l_{n}$. In this work, this requirement, written as
$l_{in}$\,=\,$2\lambda_{Q}(T,I_{dc},B)>>l_{n}$\,=\,$2\xi(T)$, is
satisfied.

In the process of multiple Andreev reflections, a quasi-electron
(quasi-hole) with a energy smaller than the superconducting gap,
located between two NS interfaces, is reflected as a quasi-hole
(quasi-electron) alternately from both NS interfaces until it
reaches the energy $V(I_{dc})en$\,=\,$2\Delta(T,B)$. As a result
of $n$ reflections, superconducting pairs in an amount equal to
$m$\,=\,$n/2$ or $m$\,=\,$(n/2)-1$ appear in the superconducting
region of the SNS junction or the phase slip center. Thus, the
effective superconducting charge increases by a factor of $m$. The
probability of observing MAR usually decreases with an increase in
the number of reflections $n$, but from the experiment \cite{c18}
it follows that MAR can be observed for a large number of
reflections (up to $n$\,=\,$32$). In \cite{c18}, it was found that
MAR and very strong quasiparticle heating in the core of the phase
slip center or SNS junction, formed in a quasi-one-dimensional
superconducting aluminum wire, cause an appearance of current
singularities in the form of a plateau on curves $V(I)$ at
voltages $V_{pl,n}(T,B)$\,=\,$2\Delta(T,B)/ne$, corresponding to
the subharmonics of the superconducting gap. Here $n$ is a certain
integer dependent on $T$, \,$B$ \,and \,$V$.

The ratio $2\Delta(T,B)/eV_{n}(I_{dc})\approx 146/14 \approx 10$
shows the possible average MAR number for the $V(B)$ function (the
curve\,2 of Fig.\,\ref{f3}). We assume that the maximum value of
this ratio can be greater. It is expected that the instantaneous
voltage $V(I_{dc})$ will vary from a value close to zero to a
value close to the maximum voltage
$V_{n}(I_{dc})$\,=\,$20$\,$\Omega$\,$\times (I_{dc})\approx
160-200$\,$\mu$V due to the presence of two different (less and
more dissipative) states at currents $I_{dc}$\,=\,$8-10$\,$\mu$A.
Therefore, the number of reflections can vary from one to the
maximum possible value $2\lambda_{Q}(T,I_{dc},B)/\xi(T)\approx
18-36$.

\section{CONCLUSION}
We have found that a superconducting mesoscopic aluminum ring,
pierced by a magnetic flux and biased with a direct current
$I_{dc}$ at $T$ slightly below $T_{c}$, can be in two different
(less or more dissipative) states.

The appearance of a certain state depends on the direction of the
field sweep. Quantum magnetoresistive $hc/2e$ periodic
oscillations of the Little-Parks type are observed in a less
dissipative state at $T$, corresponding to the lowest $R(T)$
transition, and low $I_{dc}$ currents. In this case, $hc/2e$
oscillations are not found in the more dissipative state and the
$V(B)$ function shows negative magnetoresistance. At lower $T$ and
higher $I_{dc}$, both parts of the $V(B)$ function, corresponding
to both dissipative states, exhibit two NMR regions (near
$B$\,=\,$0$ and in low fields). These NMR regions arise due to the
formation of a superconducting  barrier in wide current wires,
which prevents the overheated region of the structure from cooling
and non-monotonically depends on the field. Unusual quantum $V(B)$
oscillations can be observed in both NMR regions.

For the first time fractional $(hc/2e)/m$ (where $m=2-20$)
periodic $V(B)$ oscillations have been studied in low fields.
These oscillations were not described theoretically and measured
previously. The decrease in the oscillation period by a factor of
$m$ can be interpreted as an increase in the effective charge of
the Cooper pairs by a factor of $m$, which occurs as a result of
multiple Andreev reflections in the phase slip center or SNS
junction formed in the ring.

The functions $V(B)$, measured in other structures, also show two
dissipative states, the negative magnetoresistance and fractional
$(hc/2e)/m$ periodic oscillations. At the same time, the amplitude
corresponding to the fundamental oscillation period greatly
exceeded the amplitudes corresponding to fractional periods.

\section{ACKNOWLEDGMENTS}
This work was supported by the Ministry of Education and Science
of the Russian Federation in the framework of State Assignment
\#075-00475-19-00. The authors thank V. Tulin, D. Vodolazov, V.
Lukichev, A. Melnikov for the discussions and S. Dubonos for
making structures.

\end{document}